\documentclass[runningheads]{llncs}
\usepackage{multirow}
\usepackage{graphicx}
\usepackage{overpic}
\usepackage{amsmath}
\usepackage{color,xcolor,colortbl}
\usepackage{hyperref}
\usepackage[misc]{ifsym}
\definecolor{linkcolor}{rgb}{0.0,0.0,1.0}
\hypersetup{
	colorlinks=true,
  allcolors=linkcolor,
}

\definecolor{hicell}{HTML}{FFCDCD} 
\definecolor{redf}{HTML}{FF0000} 
\definecolor{pteseg}{rgb}{0.196,0.804,0.196} 
\definecolor{necseg}{rgb}{0.733, 0.733, 0.0} 
\definecolor{encseg}{rgb}{1.0,0.0,0.0} 

\def\ie{\emph{i.e.\,}}
\def\etal{\emph{et al.\,}}
\def\textplus{\textbf{\texttt{+}}}
\def\hicell{\cellcolor{hicell}}
\newcommand{\redbf}[1]{\textcolor{redf}{\bf#1}}

\begin{document}
\title{
	E\textsubscript{1}D\textsubscript{3} U-Net for Brain Tumor Segmentation: Submission to the RSNA-ASNR-MICCAI BraTS 2021 challenge
	\thanks{
		This work was supported by a grant from the Higher Education Commission of Pakistan as part of the National Center of Big Data and Cloud Computing and the Clinical and Translational Imaging Lab at LUMS.
	}
}
\titlerunning{E\textsubscript{1}D\textsubscript{3} U-Net for Brain Tumor Segmentation}

\author{
	Syed Talha Bukhari\inst{1} \and
	Hassan Mohy-ud-Din, PhD\inst{1}\textsuperscript{(\mbox{\scriptsize{\Letter}})}
}

\authorrunning{S. T. Bukhari \etal}

\institute{
	\textsuperscript{1}Department of Electrical Engineering,
	Syed Babar Ali School of Science and Engineering,
	LUMS,
	54792,
	Lahore, Pakistan\\
	\email{hassan.mohyuddin@lums.edu.pk}
}

\maketitle

\begin{abstract}
Convolutional Neural Networks (CNNs) have demonstrated state-of-the-art performance in medical image segmentation tasks.
A common feature in most top-performing CNNs is an encoder-decoder architecture inspired by the U-Net.
For multi-region brain tumor segmentation, 3D U-Net architecture and its variants provide the most competitive segmentation performances.
In this work, we propose an interesting extension of the standard 3D U-Net architecture, specialized for brain tumor segmentation.
The proposed network, called E\textsubscript{1}D\textsubscript{3} U-Net, is a one-encoder, three-decoder fully-convolutional neural network architecture where each decoder segments one of the hierarchical regions of interest:
whole tumor, tumor core, and enhancing core.
On the BraTS 2018 validation (unseen) dataset, E\textsubscript{1}D\textsubscript{3} U-Net demonstrates single-prediction performance comparable with most state-of-the-art networks in brain tumor segmentation, with reasonable computational requirements and without ensembling.
As a submission to the RSNA-ASNR-MICCAI BraTS 2021 challenge, we also evaluate our proposal on the BraTS 2021 dataset.
E\textsubscript{1}D\textsubscript{3} U-Net showcases the flexibility in the standard 3D U-Net architecture which we exploit for the task of brain tumor segmentation.

\end{abstract}

\section{Introduction}

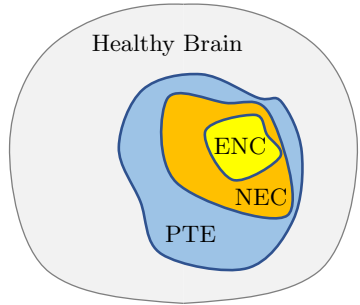
\begin{figure}[!t]
  \centering
	\begin{overpic}[clip,trim=0.7cm 1cm 0.7cm 1.1cm,width=0.4\columnwidth]{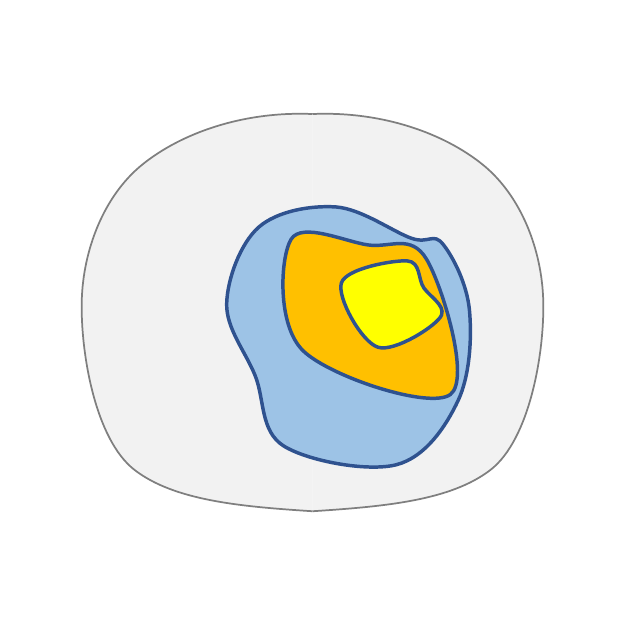}
		\put(25,72){Healthy Brain}
		\put(45,20){PTE}
		\put(64,30){NEC}
		\put(58.3,44){ENC}
	\end{overpic}
	\caption{\label{fig:braintumor}
	The brain tumor region is usually considered as a hierarchical combination of three sub-regions:
	peritumoral edema (PTE), non-enhancing core (NEC), and enhancing core (ENC) \cite{ref_brats_menze}.
	The sub-regions are schematically shown here.
	}
\end{figure}

Accurate segmentation of brain tumor sub-regions is essential in the quantification of lesion burden, providing insight into the functional outcome of patients.
In this regard, 3D multi-parametric magnetic resonance imaging (3D mpMRI) is widely used for non-invasive visualization and analysis of brain tumors.
Different MRI sequences (such as T1, T1ce, T2, and FLAIR) are often used to provide complementary information about different brain tumor sub-regions.
The brain tumor region is usually categorized into three sub-regions:
peritumoral edema (PTE), non-enhancing core (NEC), and enhancing core (ENC) \cite{ref_brats_menze}, cf. Figure \ref{fig:braintumor}.
Alternatively, these sub-regions are usually considered in a hierarchical combination:
Whole Tumor (WT: $\mathrm{PTE} \cup \mathrm{NEC} \cup \mathrm{ENC}$), Tumor Core (TC: $\mathrm{NEC} \cup \mathrm{ENC}$), and Enhancing Core (EN or ENC).

In the past decade, convolutional neural networks (CNNs) have achieved state-of-the-art performance in challenging medical image segmentation tasks.
Among various CNN architectures, the U-Net \cite{ref_ronneberger} and its variants \cite{ref_3dunet,ref_unet_dong,ref_unet_liver,ref_unet_pancreas} stand out as the most promising architectures for medical image segmentation.
However, segmentation of brain tumor and its sub-regions is challenging, even for deep neural networks, due to a number of reasons, including:
(1) Scarcity of high quality imaging data,
(2) presence of artifacts,
(3) high class imbalance,
and (4) large computational and memory requirements due to the volumetric nature of the data and its processing requirements when passed through the neural network.

In this paper, we presented an architecture comprising an encoder followed by three independent binary-output decoders (hence the name E\textsubscript{1}D\textsubscript{3} U-Net), and fused the binary segmentations through standard image-processing techniques to generate a multi-class segmentation map.
We made use of a reasonable computational budget to achieve competitive segmentation performance on the BraTS 2018 validation dataset, which we opted for since an extensive comparison with the state-of-the-art methods is readily available.
Furthermore, as a submission to the RSNA-ASNR-MICCAI BraTS 2021 challenge, we also evaluated our proposal on the BraTS 2021 dataset.

\section{Related Works}

Previous work on brain tumor segmentation poses the problem from different perspectives:
Pereira \etal \cite{ref_pereira} performed pixel-wise classification on small 2D segments through two slightly different 2D networks, one each for LGGs and HGGs.
Kamnitsas \etal \cite{ref_deepmedic} performed segmentation on 3D segments through an efficient multi-scale processing architecture, post-processed by a 3D Conditional Random Field.
Wang \etal \cite{ref_anisotropic_unc} capitalized on the hierarchical structure of tumor sub-regions by using a hierarchical cascaded of networks: one for each sub-region.
They utilized anisotropic convolutions and trained three such cascades, one for each view (axial, coronal, sagittal).
Thus, the overall architecture requires 9 trained 2.5D networks to generate a single prediction.
Dong \etal \cite{ref_unet_dong} used a 2D U-Net to segment each 3D mpMRI volume in slices.
The method is fast in training and testing and has fewer computational requirements, but is massively over-parameterized ($\approx35$ million parameters) and does not capitalize on the 3D contextual information.
Isensee \etal \cite{ref_nonewnet} used an ensemble of multiple 3D U-Nets trained on a large dataset, and focused on minor improvements to provide competitive segmentation performance.
Myronenko \cite{ref_myronenko} proposed an encoder-decoder architecture with an additional input reconstruction branch that guides and regularizes the encoder.
The network stands out in terms of segmentation performance but is not implementable in a reasonable computational budget (the author mentions 32GB of GPU memory).
Xu \etal \cite{ref_cascade_xu} used an architecture composed of a common feature extractor which branches out to an attention-guided cascade of three relatively smaller 3D U-Nets to segment each hierarchical tumor sub-region.
Each U-Net contains feature bridge modules, and the cascade is coupled by attention blocks to achieve a competitive segmentation performance.

Our proposed framework is independently developed from, but similar in essence to, the very recent work by Daza \etal \cite{ref_daza_cerberus}.
The authors used a one-encoder, four-decoder architecture where three decoders perform binary segmentation (one for each hierarchical tumor sub-region) and the fourth decoder (arising from a \emph{learned} linear combination of the learned parameters of the three binary decoders) performs the effective multi-class segmentation.

\section{Methodology}

\subsection{E\textsubscript{1}D\textsubscript{3} U-Net: One Encoder, Three Decoders}\label{ssec:e1d3}

\begin{figure}[tbp]
  \includegraphics[width=\columnwidth]{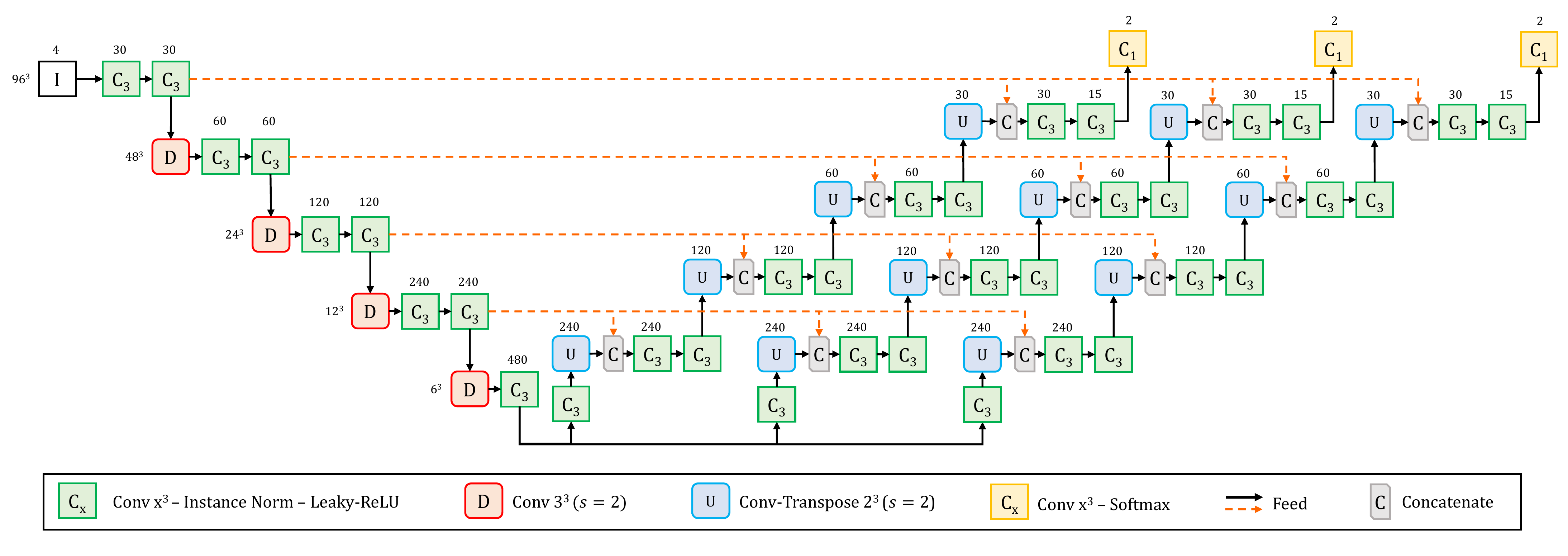}

	\caption{\label{fig:architecture}
		The proposed E\textsubscript{1}D\textsubscript{3} U-Net architecture is portrayed here.
		Each block denotes a \emph{layer}, with the output spatial dimensions and number of feature maps noted beside it.
		The first layer \framebox{I} denotes the input.
		All convolutional layers use zero-padding of $k-1$ (where $k$ is the kernel-size) on both ends of each spatial dimension.
		Strided \textit{Conv} layers (stride denoted by $s$) are used to down-sample the feature maps, while strided \textit{Conv-Transpose} layers are used to up-sample the feature maps.
		\textit{Copy and Concatenate} operation concatenates input feature maps with the corresponding output from the appropriate \textit{Conv-Transpose} layer.
		Leaky-ReLU activation uses a \emph{leakyness} parameter value of $0.01$.
	}
\end{figure}

The \emph{baseline} network in our study was based on the 3D No New-Net architecture \cite{ref_nonewnet} where we replaced max-pooling and tri-linear up-sampling layers with convolution-based up/down-sampling (as recommended in \cite{ref_nnunet_brats}).
We refer to this baseline architecture as E\textsubscript{1}D\textsubscript{1} U-Net, which is a variant of the original 3D U-Net \cite{ref_3dunet}, a fully-convolutional neural network consisting of a contracting path (encoder) and an expanding path (decoder).
The encoder performs feature extraction through successive convolutions at different levels, and the decoder combines the encoded features with the semantic information at each level to produce the output segmentation map.
Our proposed architecture, cf. Figure \ref{fig:architecture}, extends the baseline encoder-decoder architecture via a simple modification:
Adding two additional decoders, similar in design to the original decoder.
The resultant architecture consists of one encoder and three decoders, where each decoder independently receives feature maps from the encoder to generate a segmentation at the output.
We can write the overall functionality as:
\vspace{-4mm}

\begin{eqnarray}\label{eq:e1d3}
\mathbf{z} &=& (\mathbf{z}_1, \mathbf{z}_2, \mathbf{z}_3, \mathbf{z}_4, \mathbf{z}_5) \;=\; E(\mathbf{x})
\\
\mathbf{\hat{y}}_r &=& D_r(\mathbf{z}), \;\;\; r \in \{\textrm{WT}, \textrm{TC}, \textrm{EN}\}
\end{eqnarray}

\noindent where $E(.)$ and $D(.)$ respectively denote the Encoder and Decoder, $\mathbf{x}$ denotes the input sample/mini-batch, $\mathbf{z}$ is a tuple of feature maps obtained from each \emph{level} of the encoder, and $\mathbf{\hat{y}}_r$ is the output binary mask of sub-region $r$ from the decoder $D_r$.
Note that E\textsubscript{1}D\textsubscript{1} (standard U-Net) would simply be: $\mathbf{\hat{y}} = D(E(\mathbf{x}))$.
The binary segmentation maps are fused to generate the final segmentation, cf. Section \ref{ssec:testing}.
In our proposed approach, we take inspiration from the following concepts:

\begin{enumerate}
	\item \emph{TreeNets} \cite{ref_m_heads}:
	In these architectures, the network consists of multiple pathways that branch-off a common \emph{stem}.
	This allows the network branches to share parameters at the earlier stages (where more generic features are anticipated during learning) while each branch has the freedom to specialize in a different task.
	Furthermore, parameters in the stem receive accumulated supervision from multiple sources (one per branch) which may favor learning robust low-level representations.
	\item \emph{Region-based Prediction} \cite{ref_anisotropic_unc}:
	This concept proposes to organize the network in a way that it learns to optimize the hierarchical tumor regions, in contrast with segmenting each class independently.
	Such a configuration aims at directly optimizing the regions for which segmentation metrics are computed.
	In our configuration, we let each decoder specialize in one of the three hierarchical tumor sub-regions (WT, TC, and EN) by computing the loss at its output using the ground truth of corresponding sub-region (cf. Section \ref{ssec:training}).
\end{enumerate}

The network takes as input a multi-modal segment of size $96^3$ to produce an output of the same size.
The input/output size is kept relatively small to balance out the computational cost incurred by adding two additional decoders.
We noted that using a smaller input size and more feature maps per layer performs better than using a larger input size and fewer feature maps per layer, under similar settings (GPU memory, batch size).
In the latter case, a drop in performance is observed, more noticeably for TC and EN tumor sub-regions.
Note that the architecture is still very simple and does not include many of the widely used components such as residual connections and deep supervision, which may significantly increase the memory requirements.

\subsection{Training}\label{ssec:training}

Input to the network is a stack of 3D segments of shape $96^3$ from each of the multi-parametric sequences.
We extracted 3D segments at random from each subject volume within the whole-brain bounding box.
Each extracted segment was subjected to \emph{distortions} (with a $50\%$ probability), which comprised of the following operations in sequence (each with a $50\%$ probability):
Random flipping along each axis, random affine transformation, random elastic deformation, and random gamma correction.
We used a batch size of $2$, the maximum allowable in our setup.

Parameters of all convolutional layers of the networks were initialized with He-normal weights.
The networks were trained on the \emph{mean} of the objective functions applied to the output from each \emph{head} of the architecture.
The overall objective function is therefore $\mathcal{L} = (\mathcal{L}_{\mathrm{WT}} + \mathcal{L}_{\mathrm{TC}} + \mathcal{L}_{\mathrm{EN}})/3$, where each $\mathcal{L}_{x}$ is a non-weighted sum of the Soft Dice loss and the Cross-entropy loss functions, \ie $\mathcal{L}_{x} = -\mathrm{SoftDice}+\mathrm{CrossEntropy}$.
Stochastic Gradient Descent with Nesterov momentum ($0.99$), regularized by a weight decay of $10^{-6}$, optimized the network.
The learning rate was initially set to $\eta_0=10^{-2}$ and was modified at epoch-ends with a polynomial-decay policy $\eta_t = \eta_0(1-t/t_{max})^{0.9}$, where $\eta(t)$ denotes the learning rate at the $t$-th epoch and $t_{max}$ denotes the total number of epochs ($500$ in our setting).

\subsection{Testing}\label{ssec:testing}

\begin{figure}[tbp]
	\centering
	\begin{overpic}[clip,trim=0.5cm 0.4cm 0.2cm 0.3cm,width=0.6\columnwidth]{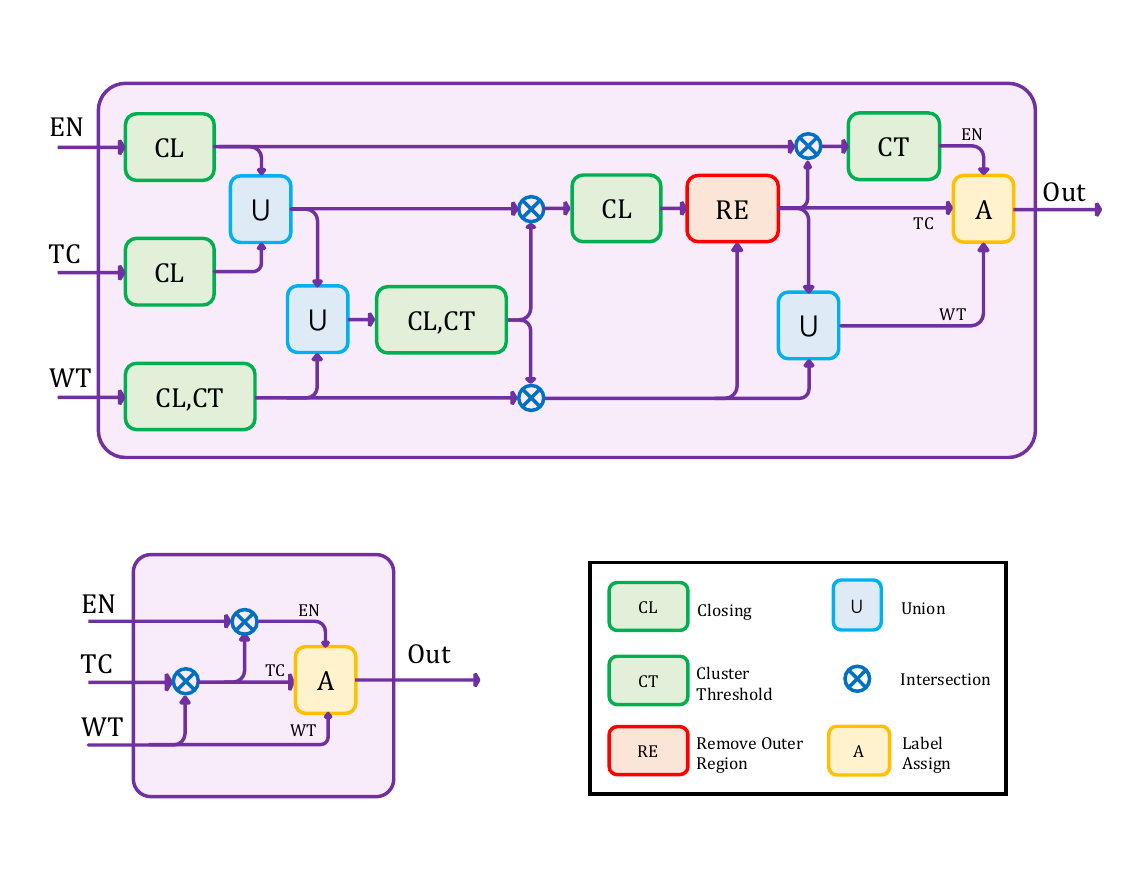}
		\put(35,31){\footnotesize Method of \cite{ref_anisotropic_unc}}
		\put(2,-1.5){\footnotesize Na\"ive Method (NvP)}
		\put(65,-1.5){\footnotesize Legend}
	\end{overpic}
	\caption{\label{fig:label_fusion}
		Label fusion procedure takes binary segmentation maps of WT, TC, EN regions and yields a multi-class segmentation map.
		\emph{RE} block uses WT and TC binary maps to remove TC region that exists outside WT region.
		Refer to the code for implementation details.
	}
\end{figure}

During inference, segments of shape $96^3$ (multi-parametric stack) were extracted from within the bounding box of the whole-brain region.
Segments were extracted with a $50\%$ overlap along each spatial axis, and \emph{softmax} outputs from the network were averaged at all regions of overlap.
The predicted hierarchical regions were fused to generate a multi-class segmentation map via a combination of morphological processing, cluster thresholding, and masking operations \cite{ref_anisotropic_unc}, cf. Figure \ref{fig:label_fusion}.
The operations are ordered to impose the following constraints: (1) The segmentations should be locally consistent and should not contain empty holes within the foreground (tumorous) region, (2) predicted tumor sub-regions in the segmentations should obey the hierarchical structure ($\mathrm{EN} \subseteq \mathrm{TC} \subseteq \mathrm{WT}$), and (3) imposition of tumor hierarchy should not result in under-segmentation of tumor sub-regions.

In addition to raw segmentation results, we also present (where mentioned) results for test-time augmentation (TTA) \cite{ref_nonewnet,ref_myronenko} in which inference is performed on the original 3D mpMRI volume and its seven additional \emph{transformed} versions.
These transformations comprised of flipping along each of the three orthogonal axes (axial, coronal, and sagittal) individually as well as in combinations.
The resulting probability maps were averaged (after un-flipping) to generate a unified probability map for each hierarchical region, before fusing the regions together to generate a multi-class segmentation map.

\section{Experiments}

\subsection{System Specifications}
For all experiments, we used open-source Python packages:
NumPy \cite{ref_numpy}, NiBabel \cite{ref_nibabel}, PyTorch \cite{ref_pytorch}, and TorchIO \cite{ref_torchio}.
We used a computer system with 64GB RAM and an NVIDIA RTX 2080Ti (11 GB) GPU.
The source code for our method is made publicly available\footnote{\url{https://github.com/Clinical-and-Translational-Imaging-Lab/brats-e1d3}}.

\subsection{Dataset and Preprocessing}
To demonstrate the effectiveness of our proposed architecture, we opt for the publicly available BraTS 2018 and 2021 datasets \cite{baid2021rsna,bakas2017segmentation1,bakas2017segmentation2,ref_brats_bakas,ref_brats_menze}.
The BraTS 2018 dataset consists of 285 training subjects (HGGs and LGGs) and 66 validation subjects.
The BraTS 2021 dataset consists of 1251 training subjects (HGGs and LGGs) and 219 validation subjects.
Both datasets comprise 3D mpMRI scans (including T1, T1ce, T2, and FLAIR), with the manual segmentation of tumor sub-regions (comprising peritumoral edema, non-enhancing tumor, enhancing tumor, and healthy/background region) available only for the training subjects.
For both BraTS 2018 and BraTS 2021 datasets, the training dataset was split into a training-fold and a validation-fold with a $9{:}1$ ratio.
Additionally, as a submission to the RSNA-ASNR-MICCAI BraTS 2021 challenge, we performed $5$-fold cross validation over the BraTS 2021 training subjects.
Predicted segmentations for each validation dataset are evaluated via the online portal provided by the
organizers of the BraTS challenge\footnote{CBICA Image Processing Portal; \url{https://ipp.cbica.upenn.edu/}}.
We also provide scores on the BraTS 2021 testing dataset comprising 570 subjects (data and ground truth not provided), for which we provided the challenge organizers with a containerized inference routine of our method.

Before training/testing, we normalized each 3D MRI volume independently to zero-mean and unit-variance within the whole-brain region.

\begin{table}[!th]
	\caption{\label{table:scores_brats18val}
		Comparison of the state-of-the-art methods on the BraTS 2018 validation dataset.
		Mean of the Dice similarity coefficient (\%) and $95$-th percentile Hausdorff distance (mm) for each region were computed by uploading the predicted segmentations on the online evaluation portal of the BraTS challenge.
		`$*$' indicates the methods we implemented/evaluated on our system. For ensemble methods, $(x)$ indicates an ensemble of $x$ networks.
		Best scores for each section are in \redbf{bold-face}.
	}
	\begin{center}\vspace{-5mm}
		\renewcommand{\arraystretch}{1.2}
		\resizebox{\textwidth}{!}{
			\begin{tabular}{|c|c|c|c|c|c|c|}
				\hline
				\multirow{2}{*}{\textbf{Method}} & \multicolumn{3}{c|}{\textbf{Dice (\%)}} & \multicolumn{3}{c|}{\textbf{Hausdorff-95 (mm)}} \\ \cline{2-7}
				& \textbf{WT} & \textbf{TC} & \textbf{EN} & \textbf{WT} & \textbf{TC} & \textbf{EN} \\

				\hline\hline
				\multicolumn{7}{c}{Ensemble Methods} \\
				\hline\hline
				3D VAE$^{(10)}$ \cite{ref_myronenko} & \redbf{91.0}   & \redbf{86.7}   & \redbf{82.3}   & 4.52 	& 6.85 	& 3.93 \\

				No New-Net$^{(10)}$ \cite{ref_nonewnet}	& 90.9    & 85.2    & 80.7    & 5.83 	& 7.20 	& 2.74 \\

				Kao \etal$^{(26)}$ \cite{ref_kao} & 90.5 & 81.3 & 78.8 & \redbf{4.32} & 7.56 & 3.81 \\

        Cerberus {\scriptsize\textplus TTA}$^{(5)}$ \cite{ref_daza_cerberus} & 89.5 & 83.5 & 79.7 & 7.77 & 10.30 & 4.22 \\

				Anisotropic-RegCascade$^{(3)}$ \cite{ref_anisotropic_unc} & 90.3 \scriptsize{$\pm 5.7$}    & 85.4 \scriptsize{$\pm 14.2$}    & 79.2 \scriptsize{$\pm 23.3$}   & 5.38 \scriptsize{$\pm 9.3$}	& \redbf{6.61} \scriptsize{$\pm 8.6$} 	& \redbf{3.34} \scriptsize{$\pm 4.2$} \\

				\hline\hline
				\multicolumn{7}{c}{Single-prediction Methods} \\
				\hline\hline

				3D VAE \cite{ref_myronenko} & 90.4   & \redbf{86.0}   & 81.5   & \redbf{4.48} 	& 8.28 	& 3.80 \\

				Cascaded-Attention-Net \cite{ref_cascade_xu} & \redbf{90.7} & 85.1 & 80.8 & 5.67 & \redbf{6.02} & 3.00 \\

				Cascaded V-Net \cite{ref_cascade_vnet} & 90.5 & 83.6 & 77.7 & 5.18 & 6.28 & 3.51 \\

				3D-SE-Inception \cite{ref_3dseincepresnet} & 90.1 & 81.3 & 79.8 & 6.37 & 8.84 & 4.16 \\

				Cross-Modality GAN \cite{ref_gan_brats} & 90.3 & 83.6 & 79.1 & 5.00 & 6.37 & 3.99 \\

				HDC-Net \cite{ref_hdcnet} & 89.7 & 84.7 & 80.9 & 4.62 & 6.12 & 2.43 \\

				OMNet \cite{ref_omnet} & 90.4 & 83.4 & 79.4 & 6.52 & 7.20 & 3.10 \\

				\hline\hline
				\multicolumn{7}{c}{Proposed Method \& Ablation Studies} \\
				\hline\hline

				\hicell E\textsubscript{1}D\textsubscript{3}$^{*}$ & 91.0 \scriptsize{$\pm 5.4$} & \redbf{86.0} \scriptsize{$\pm 15.5$} & 80.2 \scriptsize{$\pm 22.9$} & 6.56 \scriptsize{$\pm 12.8$} & 5.06 \scriptsize{$\pm 6.8$} & 3.02 \scriptsize{$\pm 4.2$} \\

				\hicell E\textsubscript{1}D\textsubscript{3} {\scriptsize\textplus TTA}$^{*}$ & \redbf{91.2} \scriptsize{$\pm 5.6$}   & 85.7 \scriptsize{$\pm 16.6$} & \redbf{80.7} \scriptsize{$\pm 21.8$} & 6.11 \scriptsize{$\pm 12.1$} & 5.54 \scriptsize{$\pm 7.5$} & 3.12 \scriptsize{$\pm 4.1$} \\

				E\textsubscript{1}D\textsubscript{1} (Baseline)$^{*}$ & 90.5 \scriptsize{$\pm 6.4$} & 84.0 \scriptsize{$\pm 18.8$} & 77.6 \scriptsize{$\pm 25.6$} & 6.44 \scriptsize{$\pm 12.8$} & \redbf{5.04} \scriptsize{$\pm 6.0$} & 3.67 \scriptsize{$\pm 6.6$} \\

				E\textsubscript{1}D\textsubscript{1} (Baseline) {\scriptsize\textplus TTA}$^{*}$ & 90.8 \scriptsize{$\pm 6.0$} & 83.3 \scriptsize{$\pm 20.6$} & 78.4 \scriptsize{$\pm 24.6$} & \redbf{5.38} \scriptsize{$\pm 9.8$} & 6.13 \scriptsize{$\pm 8.8$} & 3.38 \scriptsize{$\pm 6.2$} \\

        E\textsubscript{1}D\textsubscript{1}-Wide$^{*}$ & 89.6 \scriptsize{$\pm 9.8$} & 83.7 \scriptsize{$\pm 19.4$} & 77.7 \scriptsize{$\pm 25.8$} & 6.38 \scriptsize{$\pm 12.2$} & 6.02 \scriptsize{$\pm 7.8$} & 3.78 \scriptsize{$\pm 7.6$} \\

				E\textsubscript{1}D\textsubscript{3}-Br$^{*}$ & 90.8 \scriptsize{$\pm 6.0$} & 85.4 \scriptsize{$\pm 16.0$} & 80.0 \scriptsize{$\pm 22.1$} & 7.02 \scriptsize{$\pm 13.3$} & 5.36 \scriptsize{$\pm 5.9$} & 3.13 \scriptsize{$\pm 3.7$} \\

				E\textsubscript{1}D\textsubscript{3}-Ens$^{*}$ & 90.5 \scriptsize{$\pm 6.4$} & 84.0 \scriptsize{$\pm 19.2$} & 78.7 \scriptsize{$\pm 24.5$} & 6.10 \scriptsize{$\pm 12.0$} & 5.81 \scriptsize{$\pm 6.5$} & \redbf{2.75} \scriptsize{$\pm 3.8$} \\

				E\textsubscript{1}D\textsubscript{3}-NvP$^{*}$ & 90.9 \scriptsize{$\pm 5.6$} & 85.8 \scriptsize{$\pm 15.7$} & 79.0 \scriptsize{$\pm 25.0$} & 6.83 \scriptsize{$\pm 14.4$} & 7.45 \scriptsize{$\pm 15.6$} & 3.09 \scriptsize{$\pm 4.9$} \\

				\hline
			\end{tabular}
		}
	\end{center}
\end{table}

\subsection{Segmentation Results}

\subsubsection{BraTS 2018:}

Evaluation results on the BraTS 2018 validation dataset are shown in Table \ref{table:scores_brats18val}.
In terms of DSC, E\textsubscript{1}D\textsubscript{3} (with as well as without TTA) performs competitively for the WT and TC regions, and outperforms most methods in the EN region.
Coupled with test-time augmentation, E\textsubscript{1}D\textsubscript{3} outperforms the best-performing ensemble of 3D VAE \cite{ref_myronenko} in the whole tumor region, with only a fraction of the computational cost.
E\textsubscript{1}D\textsubscript{3} with single-prediction (without TTA) performs competitively with the ten-network ensemble of No New-Net \cite{ref_nonewnet}.
These metrics show the efficacy of the proposed multi-decoder modification to the U-Net architecture, obviating the need for ensembles to obtain competitive performance.
It must be noted that the No New-Net \cite{ref_nonewnet} architecture ensemble was trained on a larger training dataset (which the authors refer to as \emph{co-training}) whereas we only make use of the BraTS 2018 training dataset.
3D VAE architecture and No New-Net architecture respectively bagged the top two positions in the BraTS 2018 challenge.
The Anisotropic-RegCascade \cite{ref_anisotropic_unc} uses a hierarchical cascade of three networks, one for each of the three tumor regions, and ensembles three different cascades, one trained for each 3D view.
E\textsubscript{1}D\textsubscript{3}, with one trained network, outperformed the hierarchical cascaded networks in all three regions, in terms of DSC.
The tumor core HD score achieved by E\textsubscript{1}D\textsubscript{3} is better than all single-prediction and ensemble methods shown in Table \ref{table:scores_brats18val}.

Since segmentation of the three hierarchical regions is not an independent task, we compare our E\textsubscript{1}D\textsubscript{3} U-Net (with independent decoders) with a variant where the decoder for tumor core region branches-off the decoder for whole tumor region (after the first up-sampling stage), and the decoder for enhancing core region branches-off the decoder for tumor core region (also, after the first up-sampling stage).
We refer to this variant as E\textsubscript{1}D\textsubscript{3}-Br.
E\textsubscript{1}D\textsubscript{3} performs slightly better than E\textsubscript{1}D\textsubscript{3}-Br and, therefore, advocates the use of three completely independent paths for WT, TC, and EN regions.
One may also attribute the improvement in performance of E\textsubscript{1}D\textsubscript{3} to greater expressivity arising from additional number of parameters added by two additional decoders.
We therefore also compared E\textsubscript{1}D\textsubscript{3} with E\textsubscript{1}D\textsubscript{1}-Wide, where the feature maps per layer were increased to match the parameter count of E\textsubscript{1}D\textsubscript{3}, and observed that this is not the case.
To emphasize on the importance of specializing each decoder, we also trained E\textsubscript{1}D\textsubscript{3}-Ens, which is similar to E\textsubscript{1}D\textsubscript{3} but with each decoder output being a multi-class probability map, which is averaged to generate the final prediction.
In this case, we see slightly worse scores for WT region but larger differences in TC and EN sub-regions.
Nevertheless, E\textsubscript{1}D\textsubscript{3}-Ens performs better overall compared to E\textsubscript{1}D\textsubscript{1} (Baseline) and E\textsubscript{1}D\textsubscript{1}-Wide, reaffirming our intuition of TreeNets.

\begin{table}[!th]
	\caption{\label{table:scores_brats21}
		Results for cross-validation on the BraTS 2021 training dataset (seen) and for evaluation on the BraTS 2021 validation and testing datasets (unseen) are presented.
		For the validation dataset, mean of the Dice similarity coefficient (\%) and $95$-th percentile Hausdorff distance (mm) for each tumor sub-region were computed by uploading the predicted segmentations on the online evaluation portal of the BraTS challenge.
		For the testing dataset, we uploaded a containerized inference routine to the online evaluation portal, which generated the segmentations and computed the corresponding metrics.
		For ensemble methods, $(x)$ indicates an ensemble of $x$ networks.
		Best scores for each section are in \redbf{bold-face}.
	}
	\begin{center}\vspace{-5mm}
		\renewcommand{\arraystretch}{1.2}
		\resizebox{\textwidth}{!}{
			\begin{tabular}{|c|c|c|c|c|c|c|}
				\hline
				\multirow{2}{*}{\textbf{Method}} & \multicolumn{3}{c|}{\textbf{Dice (\%)}} & \multicolumn{3}{c|}{\textbf{Hausdorff-95 (mm)}} \\ \cline{2-7}
				& \textbf{WT} & \textbf{TC} & \textbf{EN} & \textbf{WT} & \textbf{TC} & \textbf{EN} \\

				\hline\hline
				\multicolumn{7}{c}{Training Dataset (cross-validation)} \\
				\hline\hline

				E\textsubscript{1}D\textsubscript{3} & 92.5 \scriptsize{$\pm 9.6$} & 89.8 \scriptsize{$\pm 17.1$} & 85.6 \scriptsize{$\pm 19.4$} & 5.28 \scriptsize{$\pm 10.3$} & 4.21 \scriptsize{$\pm 9.3$} & 3.44 \scriptsize{$\pm 7.6$} \\

				\hline\hline
				\multicolumn{7}{c}{Validation Dataset (online)} \\
				\hline\hline

				E\textsubscript{1}D\textsubscript{3} (best fold) & 91.9 \scriptsize{$\pm 7.7$} & 86.5 \scriptsize{$\pm 19.1$} & 82.0 \scriptsize{$\pm 23.4$} & 4.13 \scriptsize{$\pm 5.6$} & \redbf{7.51} \scriptsize{$\pm 35.5$} & \redbf{16.61} \scriptsize{$\pm 69.6$} \\

				E\textsubscript{1}D\textsubscript{3}{\scriptsize\textplus TTA} (best fold) & 92.3 \scriptsize{$\pm 7.5$} & \redbf{86.6} \scriptsize{$\pm 20.1$} & \redbf{82.6} \scriptsize{$\pm 22.9$} & \redbf{3.99} \scriptsize{$\pm 5.7$} & 8.23 \scriptsize{$\pm 36.2$} & 18.14 \scriptsize{$\pm 73.6$} \\

				E\textsubscript{1}D\textsubscript{3}$^{(5)}$ & 92.3 \scriptsize{$\pm 7.8$} & 86.3 \scriptsize{$\pm 21.0$} & 81.8 \scriptsize{$\pm 23.7$} & 4.34 \scriptsize{$\pm 6.5$} & 9.62 \scriptsize{$\pm 43.7$} & 18.24 \scriptsize{$\pm 73.6$} \\

				E\textsubscript{1}D\textsubscript{3}{\scriptsize\textplus TTA}$^{(5)}$ & \redbf{92.4} \scriptsize{$\pm 7.6$} & 86.5 \scriptsize{$\pm 20.8$} & 82.2 \scriptsize{$\pm 23.4$} & 4.23 \scriptsize{$\pm 6.4$} & 9.61 \scriptsize{$\pm 43.7$} & 19.73 \scriptsize{$\pm 77.4$} \\
				
				\hline\hline
				\multicolumn{7}{c}{Testing Dataset (online)} \\
				\hline\hline
				
				E\textsubscript{1}D\textsubscript{3}{\scriptsize\textplus TTA} (best fold) & 91.8 \scriptsize{$\pm 10.2$} & 86.7 \scriptsize{$\pm 24.3$} & 86.5 \scriptsize{$\pm 17.2$} & 5.68 \scriptsize{$\pm 17.7$} & 17.36 \scriptsize{$\pm 68.4$} & 9.51 \scriptsize{$\pm 48.9$} \\

				\hline
			\end{tabular}
		}
	\end{center}
\end{table}

\begin{figure}[!th]
  \flushright
	\begin{overpic}[width=0.95\columnwidth]{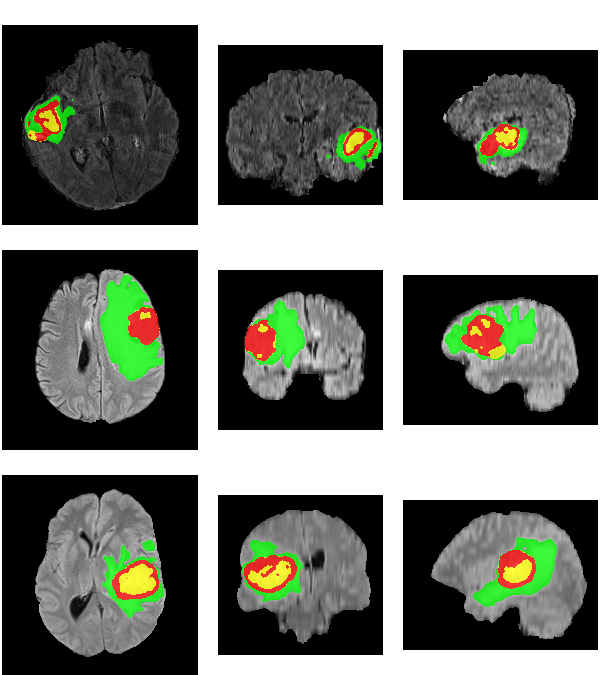}
		\put(-3, 72){\rotatebox{90}{BraTS2021\_\textbf{01691}}}
		\put(25,97.1){\underline{DSC}: 71.1, 86.9, 69.2, \underline{HD}: 15.1, 2.4, 9.9}

		\put(-3, 38){\rotatebox{90}{BraTS2021\_\textbf{01789}}}
		\put(25,63.8){\underline{DSC}: 99.0, 93.8, 86.6, \underline{HD}: 1.0, 2.0, 2.8}

		\put(-3, 5){\rotatebox{90}{BraTS2021\_\textbf{01770}}}
		\put(25,30.5){\underline{DSC}: 93.8, 96.5, 87.4, \underline{HD}: 2.0, 1.4, 2.2}

		\put(12, -2){\textbf{Axial}}
		\put(40, -2){\textbf{Coronal}}
		\put(70, -2){\textbf{Sagittal}}

	\end{overpic}

	\caption{\label{fig:segmaps_brats21val}
		Segmentations generated via E\textsubscript{1}D\textsubscript{3}{\scriptsize\textplus TTA} for 25-th percentile (top), median (middle) and 75th-percentile (bottom) subjects from the BraTS 2021 validation dataset are shown alongside metrics, evaluated by the online portal, arranged as $(WT,TC,EN)$.
		Label legend: \textcolor{pteseg}{\bf Peritumoral Edema}, \textcolor{necseg}{\bf Non-enhancing Core}, \textcolor{encseg}{\bf Enhancing Core}.
		(Ground truth is not publicly available.)
	}
\end{figure}
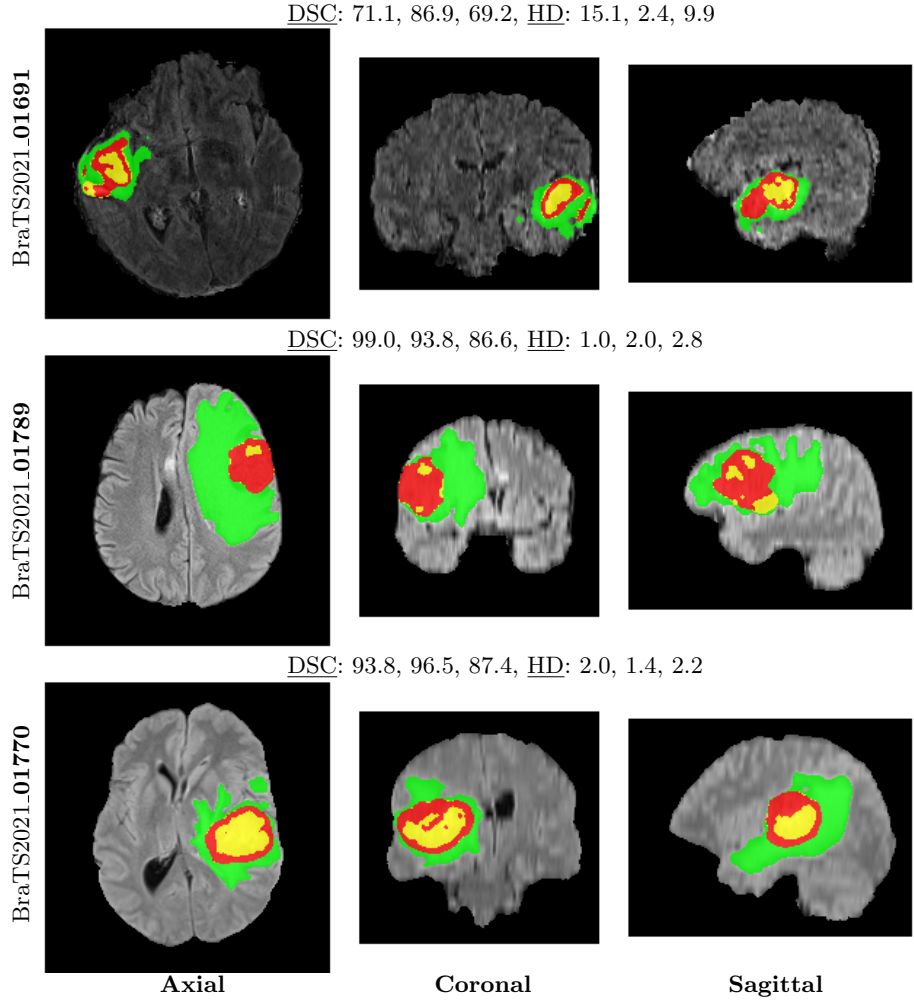

To evaluate the impact of the employed post-processing pipeline of \cite{ref_anisotropic_unc}, we use a \emph{Na\"ive} post-processing procedure, cf. Figure \ref{fig:label_fusion}, that simply imposes hierarchical constraints to generate the final segmentation map (termed as E\textsubscript{1}D\textsubscript{3}-NvP in Table \ref{table:scores_brats18val}).
We observed that the network still produces DSC and HD scores comparable to top-performing methods, emphasizing that E\textsubscript{1}D\textsubscript{3} by itself is well-designed, while the extensive post-processing method (comprising standard image-processing techniques) is recommended to yield better segmentations.
To re-emphasize, we trained and tested all architectures mentioned under the \emph{Proposed Method \& Ablation Studies} heading of Table \ref{table:scores_brats18val} using the same methodology (cf. Sections \ref{ssec:training} \& \ref{ssec:testing}), except for E\textsubscript{1}D\textsubscript{1} (\emph{training}: loss computed over single softmax output; \emph{testing}: multi-class segmentation is readily obtained) and E\textsubscript{1}D\textsubscript{3}-Ens (\emph{training}: loss averaged over each multi-class softmax output; \emph{testing}: multi-class softmax outputs are averaged to yield final prediction).
As stated previously, the difference between E\textsubscript{1}D\textsubscript{3} and E\textsubscript{1}D\textsubscript{3}-NvP is only in the post-processing pipeline used in testing.

\subsubsection{BraTS 2021:}

Results for five-fold cross-validation on the BraTS 2021 training dataset are presented along with inference results on the BraTS 2021 validation and testing datasets (unseen), cf. Table \ref{table:scores_brats21}.
E\textsubscript{1}D\textsubscript{3} attained near-peak performance with single-model predictions only, as using an ensemble of five folds did not improve significantly.
One may attribute this to a well-designed architecture which extracts rich and useful features to achieve segmentations that are hard to improve further, without significant changes.
Segmentation performance can be qualitatively judged through the segmentation maps shown in Figure \ref{fig:segmaps_brats21val}, where median, better and relatively worse cases are shown.
In the worse case, we observe an isolated island of the peritumoral edema region, which may be a slight over-segmentation causing a drop in corresponding metrics.
In the median case, the network correctly segmented a noticeably large peritumoral edema region, achieving a a DSC of $99.0$.

\section{Conclusion}

In this paper, we proposed a simple extension of the U-Net architecture specialized for brain tumor segmentation.
We couple an encoder with three independent decoders, where each decoder receives features maps directly from the common encoder and segments one of the three hierarchical tumor sub-regions:
whole tumor, tumor core, and enhancing core.
The resultant architecture, called the E\textsubscript{1}D\textsubscript{3} U-Net, provided single-model segmentation performance comparable to many state-of-the-art networks, within a reasonable computational budget and without major architectural novelties such as residual connections and deep supervision.
Through this work, we demonstrated the flexibility of the U-Net architecture, which can be exploited for the task at hand.

%
%
\bibliographystyle{splncs04}
\bibliography{references}

\end{document}